\documentclass[prb,superscriptaddress,showpacs,floatfix,twocolumn]{revtex4}
\usepackage{times}
\usepackage[dvips]{graphicx}
\usepackage{amsmath}
\usepackage{epsfig}
\usepackage{dcolumn}

\def\srvo3{SrVO$_3$}

\def\t2g{$t_\textrm{2g}$}

\newcommand{\dd}{\mathrm{d}}

\begin{document}

\title{ 
Dynamical screening effects in correlated materials:
plasmon satellites and spectral weight transfers
from a Green's function ansatz to extended 
dynamical mean field theory
}

\begin{abstract}
Dynamical screening of the Coulomb interactions
in correlated electron systems results in a
low-energy effective problem with a dynamical
Hubbard interaction $\mathcal{U}(\omega)$.
We propose a Green's function ansatz for the 
Anderson impurity problem with 
retarded interactions, in which the 
Green's function factorizes into a contribution
stemming from an effective static-U problem
and a bosonic high-energy part introducing
collective plasmon excitations.
Our approach relies on the scale separation of 
the low-energy properties, related to the instantaneous 
static $U$, from the intermediate to high energy
features originating from the retarded part of the interaction.
We argue that for correlated materials where retarded
interactions arise from downfolding higher-energy
degrees of freedom,
the characteristic frequencies are typically in the
antiadiabatic regime.
In this case, accurate approximations to the bosonic
factor are relatively easy to construct, with the
most simple being the boson factor of the dynamical
atomic limit problem. 
We benchmark the quality of our
method against numerically exact continuous time quantum Monte Carlo
results for the Anderson-Holstein model both, at half- and quarter-filling.
Furthermore we study the Mott transition within the
Hubbard-Holstein model within extended dynamical
mean field theory.
Finally, we apply our technique to a
realistic three-band Hamiltonian for \srvo3. 
We show that our approach reproduces both, the effective mass renormalization 
and the position of the lower Hubbard band by means of a 
dynamically screened $U$,
previously determined \emph{ab-initio} within the constrained 
random phase approximation.
Our approach could also be used within schemes beyond dynamical
mean field theory, opening a quite general way of describing
satellites and plasmon excitations in correlated materials.
\end{abstract}

\author{Michele Casula}
\affiliation{CNRS and Institut de Min\'eralogie et de Physique des Milieux condens\'es,
case 115, 4 place Jussieu, 75252, Paris cedex 05, France}
\affiliation{Centre de Physique Th{\'e}orique, Ecole Polytechnique, CNRS, 91128 Palaiseau
Cedex, France}
\affiliation{Japan Science and Technology Agency, CREST}
\author{Alexey Rubtsov}
\affiliation{Department of Physics, Moscow State University, 119992 Moscow, Russia}
\author{Silke Biermann}
\affiliation{Centre de Physique Th{\'e}orique, Ecole Polytechnique, CNRS, 91128 Palaiseau
Cedex, France}
\affiliation{Japan Science and Technology Agency, CREST}


\maketitle

\section{Introduction}
\label{intro}

Over the last years, significant progress has been made
in the modelization of strongly correlated materials.
Such systems typically contain partially filled
$d$ or $f$ orbitals\cite{antoine_rmp,srvo3_exp4_intro},
which lie relatively close to the nuclei.
Electronic Coulomb interactions can then induce
substantial corrections to a Bloch one-particle picture,
ranging from renormalizations of effective parameters
in the sense of Landau to full localization of the
$d$- or $f$-degrees of freedom in the Mott insulator.

One of the difficulties in describing such correlation 
effects is to separate the usually rather small 
energy range of the correlated $d$ and $f$ orbitals
from the larger energy scale of the itinerant 
degrees of freedom,
e.g. the $p$-orbitals of ligand atoms, but also of
higher or lower lying states of the transition
metal, rare earth or actinide atoms themselves.
The former orbitals are mainly responsible for 
the low-energy physical properties of the compounds,
while the latter act as a screening medium, 
setting in particular the actual value
of the Coulomb repulsion of the correlated
degrees of freedom.

As discussed in Ref.~\onlinecite{langreth},
screening is a dynamic process which leads in general to 
a frequency dependent $U=U(\omega)$. In a realistic approach 
to strongly correlated materials, $U$ can be determined
at the random phase approximation (RPA) level\cite{crpa}, 
once the bands and their eigenstates 
are computed by an ab-initio calculation.
The static value $U_0=U(0)$,
evaluated in this way\cite{crpa2_srvo3_lmto,crpa3_srvo3_wannier}, 
has been recently used in the
dynamical mean field theory (DMFT) calculations of 
materials\cite{lda_dmft_lapw},
and the effect of the frequency dependent screening has 
been either neglected or 
empirically taken into account by adjusting the effective static $U_0$.
The DMFT approach, combined with the density functional theory (DFT),
is an extremely powerful tool to treat ab-initio
strongly correlated systems, once the low-energy model is 
determined\cite{downfolding}.
However, very little is known on the impact of the frequency dependence
of the interaction in the low-energy part of the spectrum.
The hardest obstacle in order to include the dynamic $U$ into the DMFT
framework has been the lack of a reliable solver 
for the quantum impurity problem with 
a frequency dependent Hubbard interaction. 

The $U(\omega)$ computed so far for various 
materials\cite{crpa2_srvo3_lmto,crpa3_srvo3_wannier,lda_dmft_lapw} 
shows some common features. The unscreened $U$ ($U_\infty$) 
is up to an order of magnitude 
larger than the screened one ($U_0$). The frequency dependence, 
although complicated, 
can be represented roughly by a single plasmon frequency $\omega_0$, whose 
value is usually much larger than the
bandwidth (antiadiabatic plasmon). 
If one wants to deal with 
its precise form, $\mathcal{U(\omega)}$ 
can be resolved in many plasmon contributions, which
characterize the screening process.
These features make the problem difficult: 
The large $U_\infty$ rules out 
the application of traditional weak coupling expansion 
methods, while the presence of 
many plasmons prevents the direct use of methods developed in the context 
of the Hubbard-Holstein Hamiltonian
\cite{hewson_book,nrg_holstein,capone_prl,capone_prb,gutzwiller,lang_firsov_qmc,hubbard_holstein,hubbard_holstein2},
as many bosonic baths coupled to the fermion degrees of freedom 
will be necessary to fully resolve $U(\omega)$.
Recently, 
this problem has been overcome by a continuous time quantum Monte Carlo (CTQMC) 
solver proposed by Werner and Millis\cite{werner_millis_hybrid_expansion,werner_millis_hybrid_expansion_prl},
where a multi-plasmon Lang-Firsov transformation\cite{lang_firsov} is treated exactly
in the context of a hybridization expansion algorithm 
for the DMFT impurity Hamiltonian\cite{werner_millis_hubbard_holstein,werner_millis}.
Also the weak coupling CTQMC algorithm\cite{rubtsov_ctqmc_first,rubtsov_ctqmc}  
by Rubtsov can treat generic retarded interactions,
but it is limited to a not-so-large dynamic $U$ and not-so-large screening
frequencies, and therefore it becomes prohibitively
costly for realistic applications
of dynamic screening interactions in a multi-orbital context.

Another major problem still left 
(even if realiable Monte Carlo data are available)
is the possibility of computing
spectral properties, such as high-energy plasmon satellites. Indeed, 
the presence of screening modes leads usually 
to a quite complicated spectrum with a series of peaks located 
at high energies (at multiples of the plasma
frequencies). Those features are hard to get
by the usual Maximum Entropy (ME) 
methods\cite{maximum_entropy_qmc,maximum_entropy}, used to 
invert the noisy QMC data in the imaginary time domain into
the spectral properties at real frequencies. The ME methods
are quite reliable at low energy, but usually are not capable 
to deal with high frequency features. 

In this paper, we present a DMFT approach based on a 
Bose factor ansatz (BFA) for the Green's function
which is able to handle a generic $U(\omega)$ interaction in 
a strong coupling antiadiabatic regime, a typical situation
in strongly correlated materials, and provides a robust and general way 
to compute the full spectrum of the frequency dependent (retarded) $U$,
with an accuracy capable to resolve the high energy satellites.
Our method is based on 
the separation between the energy scales set by the screened value $U_0$
and treated using well established 
solvers\cite{hirsch_fye,rubtsov_ctqmc_first,werner_millis_hybrid_expansion}, 
and the dynamic part treated with various levels of approximation,
the simplest and most insightful one taken from the dynamic atomic limit.

The paper is organized as follows. 
In Sec.~\ref{ansatz} we specify
the Anderson impurity model we would like to solve and 
the Green's function ansatz 
used in our method, in Sec.~\ref{atomic} we present the
dynamic atomic limit approximation (DALA) to our approach, 
in Sec.~\ref{beyond}
we show various ways to improve upon the DALA, and 
in Sec.~\ref{overview} we describe their performances.
In Secs.~\ref{results} and \ref{srvo3} we 
report our results for a single-band lattice model at half-filling and 
a three-band model with the DFT density of state (DOS) of \srvo3, respectively. 
Finally, Sec.~\ref{conclusions} summarizes our findings.

\section{The Green's function Bose factor ansatz (BFA)}
\label{ansatz}

\subsection{General model}
We discuss here the case of a multi-orbital
Hubbard-Holstein model.
In the context of the DMFT approach, one maps the full lattice problem 
into a single-site Anderson impurity problem coupled to an effective bath.
The bath is determined self-consistently by requiring that 
the impurity Green's function
equals the on-site Green's function on the lattice.\cite{antoine_rmp} 
Therefore, computing in the most effective way the Green's function of the Anderson model (AM) 
is of key importance to have a feasible DMFT scheme. Here, we have the additional complication that 
the effect of screening makes the on-site Hubbard interaction retarded.
In the Matsubara imaginary time action formalism 
the dynamic Anderson model reads:      
\begin{widetext}
\begin{equation}
{\cal S} = - \int_0^\beta \!\!\! \dd\tau \int_0^\beta \!\!\! \dd\tau^\prime \sum_{i \sigma} 
c^\dag_{i \sigma}(\tau) {\cal G}^{-1}_{0,i \sigma}(\tau-\tau^\prime) c_{i \sigma}(\tau^\prime)  
 +  \frac{1}{2} \int_0^\beta \!\!\! \dd\tau \int_0^\beta \!\!\! \dd\tau^\prime \sum_{i \sigma j \sigma^\prime} 
\left(n_{i \sigma}(\tau) - \frac{1}{2} \right) 
U_{i \sigma,j \sigma^\prime}(\tau-\tau^\prime)  
\left(n_{j \sigma^\prime}(\tau^\prime) - \frac{1}{2} \right),
\label{s_eff}
\end{equation}
\end{widetext}
where $\beta$ is the inverse temperature, 
$U_{i \sigma,j \sigma^\prime}(\tau)$ is the screened interaction, 
$(i,\sigma)$ is the set of orbital and spin indexes, 
and $(c^\dag_{i \sigma},c_{i \sigma})$ are the creation and annihilation operators
satisfying the antisymmetric commutation relations.
${\cal G}^{-1}_{0,i \sigma}(i \omega)=i \omega + \mu - \Delta_{i \sigma}(i\omega)$ 
is the effective hopping term coupled with the bath via the hybridization $\Delta_{i \sigma}(i\omega)$,
and $\mu$ is the chemical potential. 
Let us assume that the dynamic part of $U$ is 
orbital-independent and couples only to the total charge.
\begin{equation}
\label{generic_interaction}
U_{i \sigma,j \sigma^\prime}(\tau)=U^0_{i \sigma,j \sigma^\prime} \delta(\tau) + {\bar U}(\tau), 
\end{equation}
with ${\bar U}(0)=0$. Therefore, we assume that all details of the interaction due to effective
spin and Hund's couplings are embedded in the instantaneous part of the full $U$.
Here and thereafter we are going to take the convention that the instantaneous $U^0$ is the 
static (fully screened) limit of $U$ with ${\bar U}$ being 
a \emph{repulsive} contribution.

\subsection{The Bose factor ansatz}
We are mainly interested in evaluating the Green's function $G_{ij}(\tau) = \langle {\cal T} c_j(\tau) c_i^\dag(0) \rangle$ 
and its spectral properties for the model in Eqs.~\ref{s_eff} and \ref{generic_interaction}.
We are going to rewrite it in the form:
\begin{equation}
G_{ij}(\tau)= F_{ij}(\tau) G_{0,ij}(\tau),
\label{eq:ansatz}
\end{equation}
where $G_{0,ij}(\tau)$ is the Green's function for the model in Eq.~\ref{s_eff}, but with a static
on-site repulsion, namely $U_{i \sigma,j \sigma^\prime}(\tau)=U^0_{i \sigma,j \sigma^\prime} \delta(\tau)$.
We highlight that the above factorization is defined in the \emph{time domain}, a feature 
which is borrowed from the dynamic atomic solution of the problem whose form is known analytically,
as explained in Sec.~\ref{atomic}. In that limit the Green's function assumes exactly 
the form in Eq.~\ref{eq:ansatz}, with $G_0$ the instantaneous $U_0$ atomic Green's function.
The static model is much easier than the dynamic one, since it contains only the energy scales set by the screened $U (\ll U_\infty)$
and the Kondo resonance with the bath,
and it can be solved by means of various techniques\cite{hirsch_fye,rubtsov_ctqmc_first,werner_millis_hybrid_expansion}, 
which are usually very robust and efficient in this case. 
On the other hand, $F_{ij}(\tau)$ is a Bose factor, which is a functional of ${\bar U}(\tau)$, and it is not known a priori.
However, we will present various approximations where the function  $F_{ij}(\tau)$ is derived.
It contains the information of the plasmon (or phonons) excitations, and the plasmon (or phonons) satellites.

\subsection{BFA spectral properties}
A great advantage of dealing with the Green's function ansatz in Eq.~\ref{eq:ansatz}
is the possibility to compute very accurate spectral functions
over the whole energy range, including the intermediate-high energy plasmon
satellites. Indeed, since the Bose factor $F_{ij}(\tau)$ can be estimated 
analytically by means of some approximation,
its numerical value is known at machine precision, and its spectral function
$B(\omega)$ can be obtained via a Pad\'e approximant,\cite{pade} 
in an accurate and robust way.
On the other hand, a
ME approach has to be used to find the spectral
function $A_0(\omega)$ of the static Green's function $G_{0,ij}(\omega)$.
However, this does not pose any particular problem, since
there are no high-energy features in $A_0(\omega)$, and its
energy range is set by $U_0$, where the ME is reliable
in presence of data with good statistics.\cite{maximum_entropy}
The spectral function $A(\omega)$ of the full Green's function
$G_{ij}(\tau)$ expressed as a functional of $B$ and $A_0$ reads:
\begin{equation}
A(\omega)=\int_{-\infty}^\infty \!\!\! \dd\epsilon ~ B(\epsilon) 
\frac{1+e^{-\beta\omega}}{(1 + e^{-\beta(\epsilon-\omega)})(1 - e^{-\beta\epsilon})} 
A_0(\omega-\epsilon).
\label{spectral_conv}
\end{equation}
The spectral functions obtained in this way 
are reported for instance in Fig.~\ref{spectral},
which corresponds to the Green's functions 
plotted in Fig.~\ref{benchmark}. The quality of the satellite resolution
is striking, much higher than the one usually obtained 
with ME methods\cite{maximum_entropy},
particularly at energies far away from the Fermi level.

We would like to stress that the spectral convolution in Eq.~\ref{spectral_conv} is general, and
can be used not only for the approximated Green's functions we are
going to derive in Secs.~\ref{atomic} and \ref{beyond}. For instance,
our approach to compute the spectral properties can be
applied to the Green's function obtained by means of
the algorithm in Ref.~\onlinecite{werner_millis}. Given the full
Green's function $G(\tau)$ of the dynamic impurity problem of
Eq.~\ref{s_eff}, one defines an auxiliary Green's function
$G_\textrm{aux}(\tau)$ as $G(\tau)/F_\textrm{DALA}(\tau)$, with $F_\textrm{DALA}$ taken from the
atomic limit as described in Sec.~\ref{atomic}. This is an effective way to
exploit the separation of the low energy properties, kept in
$G_\textrm{aux}$, from the high frequency features correctly reproduced
by the DALA.  At this point, one computes$A_0$, the spectral representation of $G_\textrm{aux}$, by using ME,
and $B$, the spectrum of $F_\textrm{DALA}$, by means of the Pad\'e
approximant, and evaluates the full spectral function in Eq.~\ref{spectral_conv}.

\section{The dynamic atomic limit approximation}
\label{atomic}

In order to find a way to determine $F_{ij}(\tau)$ in Eq.~\ref{eq:ansatz},
for the moment let us take into account the single-orbital symmetric case.
Thus, we can drop all the orbital and spin indexes,
and simplify considerably the notation.
By inverting Eq.~\ref{eq:ansatz}, one gets:
\begin{equation}
F(\tau)= \frac{G(\tau)}{G_0(\tau)}  \approx  \left . \left( \frac{G(\tau)}{G_0(\tau)} \right) \right|_{\Delta=0},
\label{eq:approx1}
\end{equation}
where the rightmost-hand side of the above Equation is the approximation for $F(\tau)$
taken in the dynamic atomic limit (DAL), when the hybridization $\Delta$ is zero.
It turns out that the Green's function in the DAL is analytically solvable 
by means of a Hubbard-Stratonovich transformation\cite{florens_thesis}, and
therefore $G/G_0$ is exactly known in a close analytic form, such that:
\begin{equation}
F_\textrm{DALA}(\tau) 
=
\exp\left(
\frac{1}{\beta} \sum_{n \ne 0} \frac{U(i \nu_n) - U_0}{\nu_n^2} \left( e^{i \nu_n \tau}-1 \right) \right),
\label{eq:dala}
\end{equation}
where $\nu_n=2n\pi/\beta$ are bosonic Matsubara frequencies, with $n$ relative integer.

Beside the atomic limit,
this approximation is exact in the static and the non-interacting limits (in both cases $F_\textrm{DALA}(\tau)=1$).
Notice that it retains all the non-perturbative character of $G_0(\tau)$. To have a better idea 
on the quality of this approximation, we are going to test it for 
the dynamic $U$ with a single plasmon mode $\omega_0$, which is equivalent to the
Anderson-Holstein model with $U(i \nu)= U_\infty - 2 \lambda^2 \omega_0 /(\nu^2 + \omega_0^2)$
and the electron-``phonon'' coupling given by $\lambda=\sqrt{(U_\infty-U_0) \omega_0 / 2}$. 
We use the CTQMC algorithm by Rubtsov,\cite{rubtsov_ctqmc_first,rubtsov_ctqmc} 
which can handle retarded interactions 
and yield the exact Green's function
in a weak coupling regime, to benchmark our approximation 
for the particle-hole
symmetric system with $U_0=2$, $\beta=10$, and few values of $\omega_0$ and $U_\infty$.
The energy units are expressed in terms of the half bandwidth ($D/2=1$) of 
the semicircular DOS.

\begin{figure}
\begin{center}
 \includegraphics[angle=-90,width=\columnwidth]{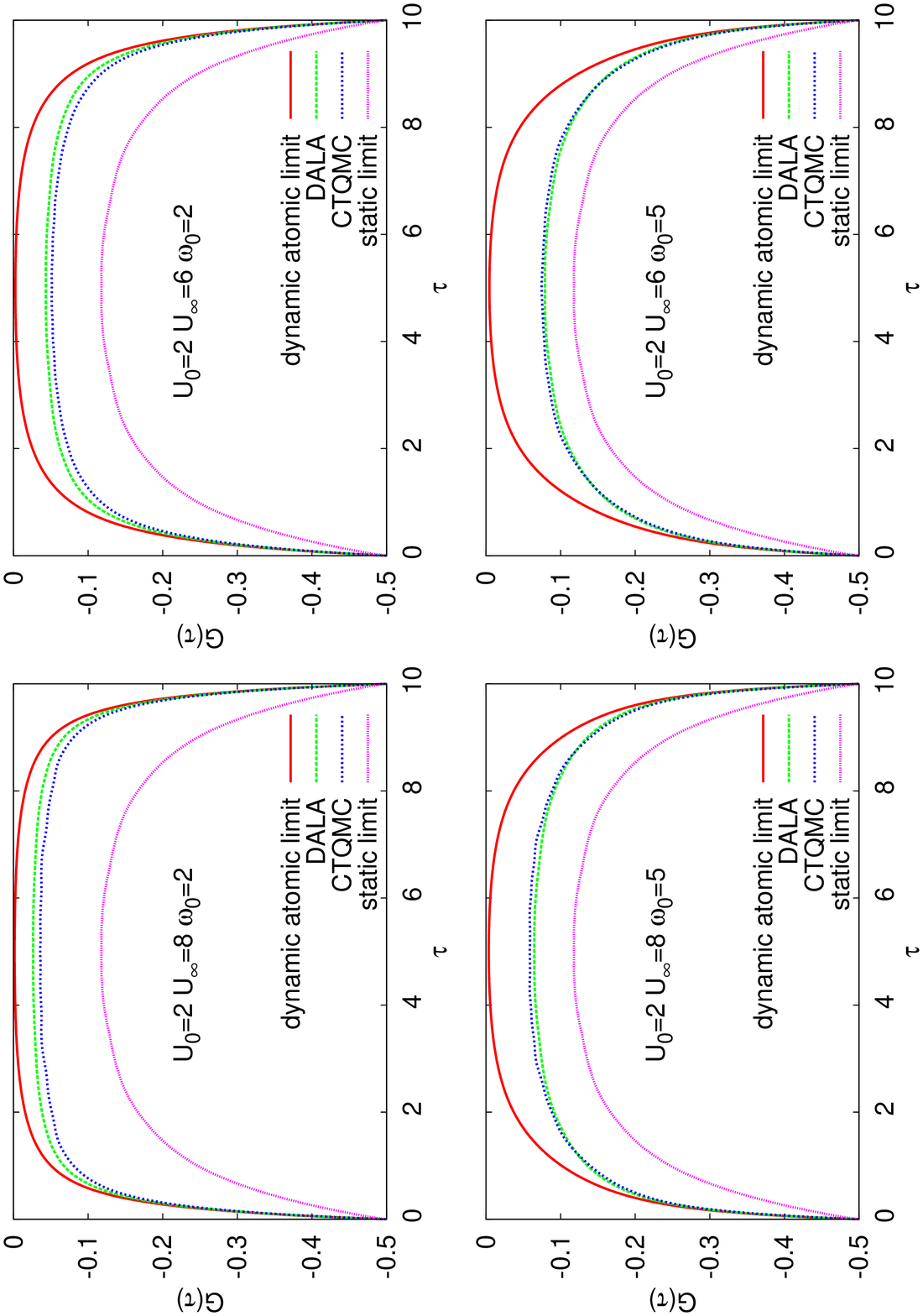}
\end{center}
\caption{\label{benchmark}
(color online) Green's functions for a half-filled Anderson model with 
dynamically screened U and a semicircular density of states at $\beta=10$.
The $G_\textrm{DALA}$ obtained by the method proposed in Sec.~\ref{atomic} is plotted (green long-dashed line)
and compared to the exact Rubtsov's CTQMC numerical result (blu dotted line). Also the dynamic atomic limit 
(red solid line) and the static one (pink dot-dashed line) are reported.
The system is computed at a quite large temperature, such that
an accurate benchmark against the numerically exact CTQMC is still possible,
even for quite large values of $U$.
 }
\end{figure}

\begin{figure}
\begin{center}
 \includegraphics[width=\columnwidth]{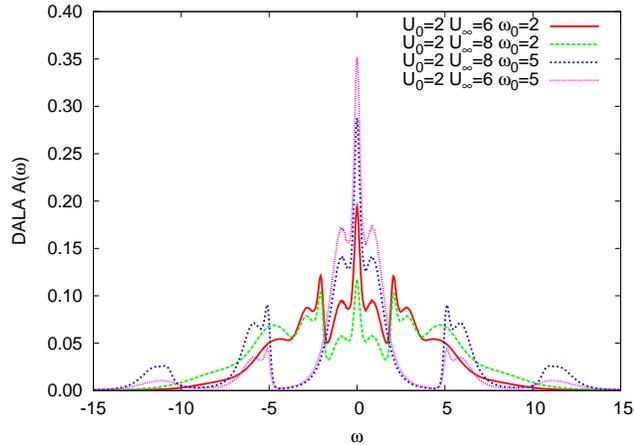}
\end{center}
\caption{\label{spectral}
(color online) Spectral function of the $G_\textrm{DALA}$ Green's functions 
reported in Fig.~\ref{benchmark}.
}
\end{figure}

As one can see from Fig.~\ref{benchmark}, the DALA works
very well for the cases analyzed, since it gives a $G_\textrm{DALA}(\tau)$ which almost
coincides with the numerically exact $G(\tau)$ given by the CTQMC algorithm.
The accuracy is particularly impressive in the case
$\beta=10$, $U_0=2$, $U_\infty=6$, and $\omega_0=5$, where the 
impact of the dynamic part is reduced by the larger $\omega_0 (\gg U_0)$,
and a smaller $U_\infty$, namely when the the energy scales of the static
part set by $U_0$ are well separated form the dynamic contributions in $U(i \nu)$.
Moreover, from Fig.~\ref{benchmark} it is apparent
that the low energy properties of the system are strongly renormalized
by the effect of the high energy components of $U$. This is quite insightful on
the importance of the dynamic screening effects in the
treatment of more realistic models that we are going to tackle in 
Sec.~\ref{results}.
This can be noted also from the spectral properties
reported in Fig.~\ref{spectral}, which
correspond to the Green's functions $G_\textrm{DALA}(\tau)$ 
plotted in Fig.~\ref{benchmark}.
A spectral weight transfer from the low frequency
spectrum to the high energy satellites is clearly visible in the Figure.

\begin{figure}
\begin{center}
 \includegraphics[width=\columnwidth]{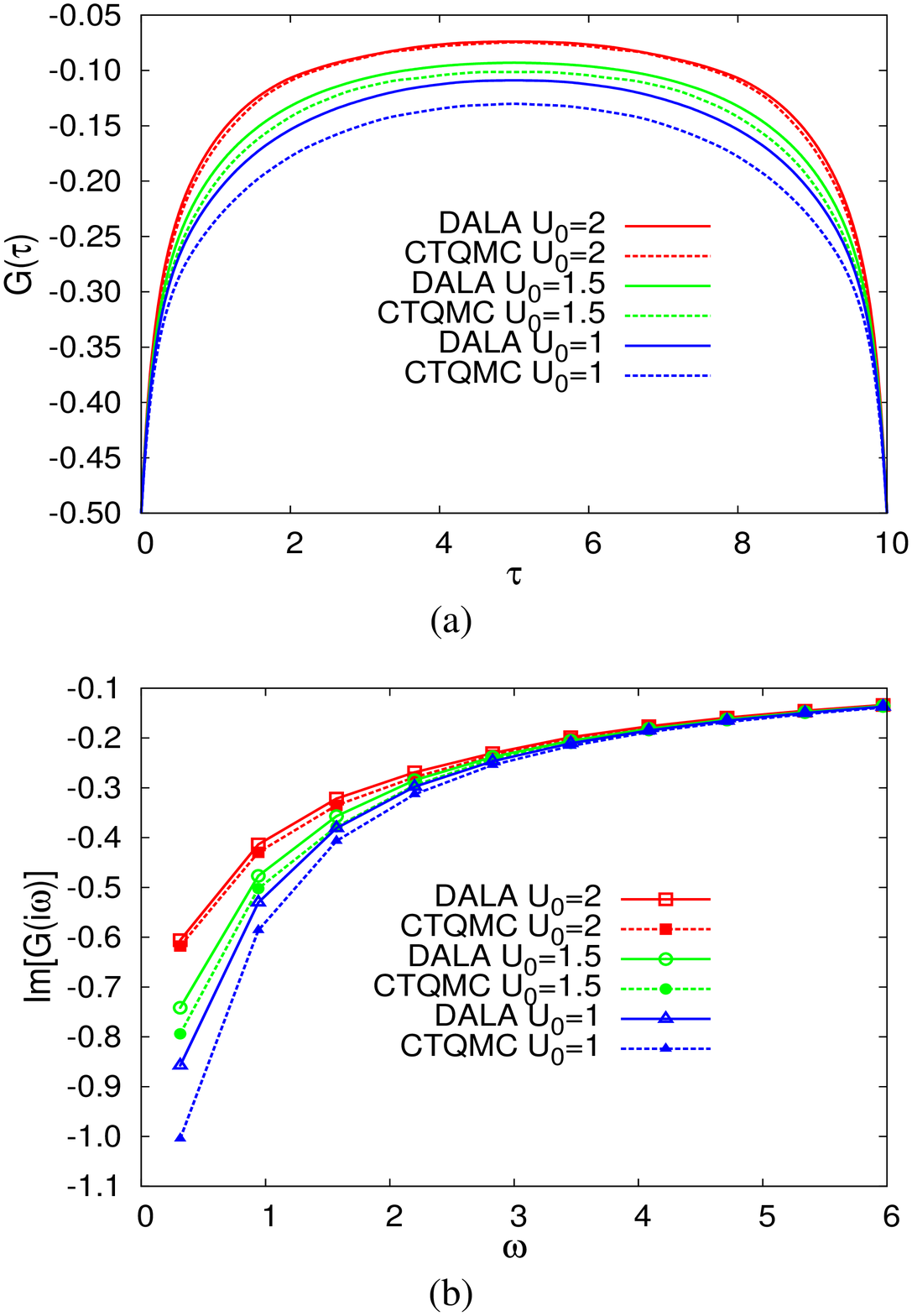}
\end{center}
\caption{\label{weaker}
(color online) Half-filled Anderson impurity
model with retarded screened interaction: 
$\omega_0=5$, $U_\infty-U_0=5$, $\beta=10$, for various values of $U_0$ 
(a) numerically exact Rubtsov's CTQMC and DALA $G(\tau)$; (b) CTQMC and DALA imaginary part of 
$G(i \omega_n)$.
}
\end{figure}

Since the DALA is obtained in the $\Delta=0$ limit, it works 
well in the intermediate-strong coupling regime, with $U_0$ and the
dynamic part large, as we have seen in the cases analyzed in
Fig.~\ref{benchmark}, where $U_0$ was quite close to the critical $U_{c2}$ 
($\approx 2.6$, see Ref.~\onlinecite{mott_transition_dmft}) for the Mott transition of the static
Hubbard model. 
However, it deteriorates as $U_0$ is getting smaller and
the hybridization $\Delta$ becomes important to set the low-energy
properties of the system.
To show this, let us take into account the Anderson-Holstein model with $U_0=2$, $\omega_0=5$, 
and $U_\infty=6$ at $\beta=10$, for which the DALA gives a result very close to
the exact one. Now, let us keep $\omega_0$ and $U_\infty - U_0$ fixed, such
that the DALA Bose factor (Eq.~\ref{eq:dala}) is unchanged,
while we vary $U_0$ from strongly to more weakly
correlated values. The resulting Green's functions are plotted in
Fig.~\ref{weaker}. It is apparent that the DALA accuracy
reduces as $U_0$ decreases, and the resulting bias is more pronounced in the 
low-energy part of the Green's function. Indeed, the discrepancy
is larger around $\beta/2$ in $G(\tau)$ (Fig.~\ref{weaker}a), which corresponds
to a larger difference at small Matsubara frequencies in $G(i \omega_n)$ (Fig.~\ref{weaker}b).
On the other hand, the high-energy tails of $G(\tau)$ are very well reproduced by the DALA,
as it is confirmed also by the inspection of $G(i \omega)$ at large $\omega_n (>4)$.

\begin{figure}
\begin{center}
 \includegraphics[angle=-90,width=\columnwidth]{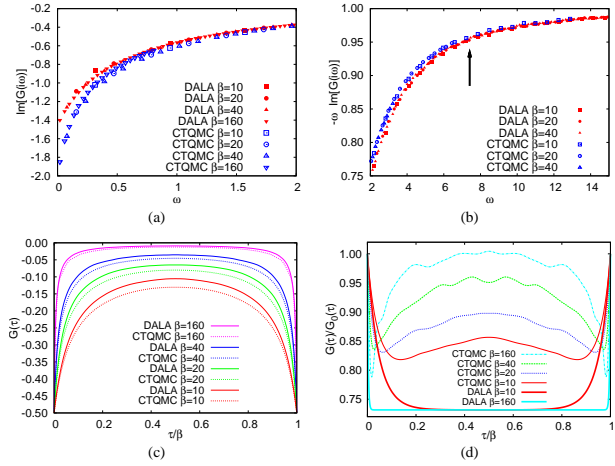}
\end{center}
\caption{\label{pinning}
(color online) Symmetric Anderson model with retarded screened interaction: 
$U_0=1.25, \omega_0=2, U_\infty=2.5$, at different $\beta$.
(a) Imaginary part of the numerically exact Rubtsov's CTQMC 
and DALA Green's functions for small Matsubara frequencies $\omega_n$;
(b) $-\omega_n \textrm{Im}[G(i\omega_n)]$ in the intermediate frequency range. The arrow indicates
the frequency when the DALA and the CTQMC Green's functions become practically indistinguishable.
The DALA reproduces correctly the intermediate-high energy behavior of the Green's function
well beyond the ``trivial'' $1/i\omega$ term;
(c) $G(\tau)$; (d) The CTQMC and DALA Bose factor $F(\tau)$. 
The thicker lines are for the DALA. The discrepancy between the
DALA and the exact factors is getting smaller as the temperature increases. The wiggles are due to
the stochastic noise of the data.
}
\end{figure}

The correct high-energy asymptotics of the DALA is a nontrivial property of this approximation,
which is borrowed from the atomic limit exactly built-in.
In order to further analyze this important feature, we take into account the 
temperature dependence of the DALA in the symmetric Anderson model with a 
dynamic interaction given by $U_0=1.25$, $\omega_0=2$, and $U_\infty=2.5$. In this
not-so-correlated case, the model can be solved exactly down to low temperatures ($\beta=160$) even
by the CTQMC algorithm, to benchmark the temperature dependence of our approximation.
Results are plotted in Fig.~\ref{pinning}. As we already found in the previous analysis, 
at large Matsubara frequencies the dependence of the Green's function is correctly given 
by the DALA, which in this case becomes almost indistinguishable 
from the exact CTQMC result for $\omega_n>7$. We note that the DALA
is capable to reproduce the decay of the imaginary part of $G(i \omega)$ well beyond the
$1/i\omega$ term, as it is apparent from Fig.~\ref{pinning}(b).
It is also worth noting that the relative accuracy of the approximation increases 
with the temperature, as it is shown in Fig.~\ref{pinning}(d). Indeed, the 
exact Bose factor $F(\tau)$ is getting closer to $F_\textrm{DALA}(\tau)$
as the temperature increases. At low temperatures, it is the $F(\beta/2)$ value which is 
poorly reproduced by the DALA. Again, this is related 
to the roughness of the approximation at low-frequency, 
which does not describe accurately the low-energy excitations 
around and below the coherent temperature. 
Indeed, the Friedel sum rule is clearly 
violated, as one can see in Fig.~\ref{pinning}(a), where the condition
$G^{''}(i0^+)=-4/D$ valid at half filling is not met by the DALA.

Therefore, going beyond the DALA is needed to capture the low-energy low-temperature features
of the spectral function, while its high-energy properties, as the plasmon satellites,
can be successfully taken into account at this level of approximation.

\begin{figure}
\begin{center}
 \includegraphics[angle=-90,width=\columnwidth]{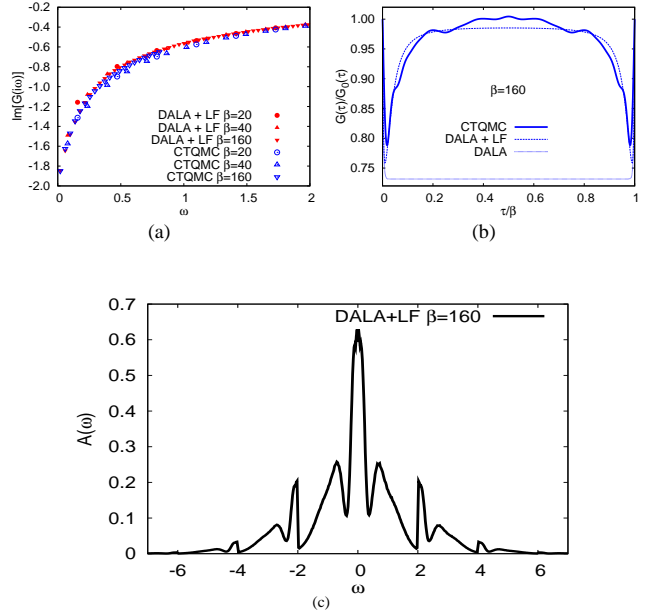}
\end{center}
\caption{\label{lang_firsov_low_T}
(color online) Symmetric Anderson model with retarded screened interaction as in Fig.~\ref{pinning}: 
$U_0=1.25, \omega_0=2, U_\infty=2.5$.
(a) Imaginary part of the numerically exact Rubtsov's CTQMC and DALA+LF Green's functions for small Matsubara frequencies $\omega_n$
at different $\beta$;
(b) The CTQMC, DALA and DALA+LF Bose factor $F(\tau)$ at $\beta=160$. 
The improvement of the $F(\tau)$ provided by the Lang Firsov correction is apparent.
The wiggles are due to the interpolation of noisy QMC data;
(c) Spectral representation of the DALA+LF Green's function in (a). 
}
\end{figure}

\section{Beyond the dynamic atomic limit}
\label{beyond}

\subsection{The DALA Lang-Firsov approximation}
\label{dalalf}

A way to improve the dynamic atomic limit in both the intermediate-low energy correlations and
low temperature regimes is provided by the Lang-Firsov approach. This approximation
has been widely used in the literature to tackle electron-phonon models
in the antiadiabatic limit, when the electron-phonon coupling is $\lambda \ll \omega_0$, 
with $\omega_0$ the phonon frequency. The same applies to models where the interaction
is retarded by the charge screening plasmons. 
In the latter case, the antiadiabatic
regime is met more often, since the plasmon frequences are larger than 
the phonon ones, usually by an order of magnitude. 
However, it should be noted that in the realistic retarded Hubbard $U$ also
the electron-plasmon coupling is stronger, as $\lambda=\sqrt{(U_{\infty}-U_0) \omega_0/2}$, and
$U_{\infty}$ is an order of magnitude larger than $U_0$. In any case, if $U_{\infty}-U_0 < \omega_0$,
the Lang-Firsov approach describes well the low energy properties 
of the system. Here, we use the Lang-Firsov
approximation in a new and original way, as a low frequency correction to the
dynamic factor of our Green's function ansatz.

The factorization in the $\tau$ space implies a convolution in the $\omega_n$ space, 
\begin{equation}
G(i \omega_n) = \frac{1}{\beta} \sum_m G_0(i \omega_m) F(i \omega_n - i \omega_m),
\end{equation}
where $F(i \nu_n)$ are the Matsubara components of the Bose factor.
A way to improve upon the DALA is to choose the $F$ factor such that the Lang-Firsov behavior 
is obtained at low frequency. This can be done, by introducing an enhanced $F$ defined as:
\begin{equation}
\label{mixing}
F_\textrm{\tiny DALA+LF}(i \nu_n) = \left \{
\begin{array}{ll}
a     & \textrm{if $\nu_n = 0$} \\
\left (1 - b \exp(-\frac{\nu_n}{c}) \right) F_\textrm{\tiny DALA}(i \nu_n)  & \textrm{elsewhere}
\end{array} 
\right.,
\end{equation}
where $a$, $b$, and $c$ are parameters determined by the following conditions:
$F(0) = F(\beta) = 1$, which keeps the correct number of particles provided by $G_\textrm{static}$;
$G(i \omega_l) = G_\textrm{LF}(i \omega_l)$ for $l=0$ (the first Matsubara frequency), 
which gives the correct Friedel sum rule
fulfilled by the Lang-Firsov approximation (and broken by the DALA);
$c$ is the crossover frequency between the LF behavior at low energy and the
atomic DALA behavior at high frequency, and its optimal value is $\approx \omega_0/10$, 
with $\omega_0$ the lowest Holstein frequency. 

The LF Green's function is given by the usual expression:
\begin{equation}
{\textstyle
G_\textrm{LF}(i \omega) = 
\frac{ \exp \left( - \frac{\lambda^2}{\omega^2_0} \right) }{ i \omega + \mu -  \lambda^2/\omega_0 -    
\exp  \left( - \frac{\lambda^2}{\omega^2_0} \right) \Delta (i \omega) - \Sigma[U_0] (i \omega)},
}
\label{LF_green_function}
\end{equation}
which clearly implies that an Anderson model with static Hubbard $U_0$ must be solved, hybridized 
via a renormalized bath $\exp  ( - \lambda^2/\omega^2_0 ) \Delta (i\omega)$. Therefore, to get
the DALA+LF Green's function one has to solve two static models (with regular and renormalized
bath) and mix them together by using our definition in Eq.~\ref{mixing}.

One should note that the Lang-Firsov Green's function in Eq.~\ref{LF_green_function} 
has been written for a single plasmon (or phonon) Anderson-Holstein model, which
gives a rough representation of the dynamically screened $U(i \nu)$ present in
\emph{ab-initio} models. For instance the c-RPA approximation for $U(i \nu)$ 
usually leads to a quite broad spectrum of screening plasmons, 
which is difficult or impossible to fit accurately by a single frequency model.
However, a generalization of the Eq.~\ref{LF_green_function} can be easily 
done by following the same lines as in the Ref.~\onlinecite{werner_millis}.
If the LF renormalization factor $\exp(-\lambda^2/\omega^2_0)$ is replaced by
\begin{equation}
\label{generalized_LF}
\exp \left( \int_0^\infty ~ \frac{\dd\tilde{\omega}}{\pi} \frac{\textrm{Im} {\bar U}(\tilde{\omega})}{\tilde{\omega}^2} \right),
\end{equation}
where ${\bar U}(\tilde{\omega}) = U(\tilde{\omega})-U_0$, and $\tilde{\omega}$ are real frequencies,
all the screening plasmons are treated on the same footing. Although the LF approximation
is accurate only for large $\tilde{\omega}$, the Eq.~\ref{generalized_LF} is a good approximation if 
$\textrm{Im}{\bar U}(\tilde{\omega})/\tilde{\omega}^2$ goes rapidly to zero for small $\tilde{\omega}$. 
For the Holstein single-mode interaction
$\textrm{Im}{\bar U}(\tilde{\omega}) = - \pi \lambda^2 (\delta(\tilde{\omega} - \omega_0) - \delta(\tilde{\omega} + \omega_0))$,
and one recovers the standard LF renormalization factor.

The improvement of the DALA+LF correction with respect to the simple DALA factor is apparent 
in Fig.~\ref{lang_firsov_low_T}(a), to be compared with Fig.~\ref{pinning}(a). 
In particular, at low frequency the LF is capable to recover 
the pinning condition ($G^{''}(i0^+) = -2$ for the semicircular density of states at half filling), 
violated by the DALA at low temperatures. This is reflected in the behavior of the
Bose factor $F_\textrm{DALA+LF}$ drawn in Fig.~\ref{lang_firsov_low_T}(b), 
whose value at $\beta/2$ goes correctly to 1 for $\beta \rightarrow \infty$, in contrast to the 
$F_\textrm{DALA}$ which does not change with temperature. Indeed, the exact $F$ factor shows 
a quite strong temperature dependence, as reported in Fig.~\ref{pinning}(d).
In Fig.~\ref{lang_firsov_low_T}(c), we plot the spectral
representation of the DALA+LF Green's function at $\beta=160$ obtained
by the DALA factor assisted analytic continuation described in Sec.~\ref{ansatz}.

\subsection{Diagrammatic first-order expansion: Gaussian cumulants (GC)}
\label{diagrammatic_fullU}

Another way to improve upon the DALA is to rely on 
the diagrammatic expansion of the interaction. This can be done in
various ways. Here we took two routes: making a cumulant 
Green's function expansion in the full $U(i \nu)$ and
the retarded part ${\bar U}(i \nu)=U(i \nu) -U_0$ only. In the former approach, 
dubbed ``Gaussian cumulants'' (GC) and described in this
Subsection, the perturbation theory is built on the
Gaussian action containing 
the hybridized $\mathcal{G}_0$ as the bare Green's function.
In the latter method, named ``instantaneous bold cumulants'' (IBC)
and introduced in Subsection \ref{diagrammatic_deltaU}, 
the perturbation expansion requires
the calculation of density-density correlators within the 
static $U_0$ model by means of the CTQMC algorithm (or other 
algorithms suitable for static interacting models).\cite{antoine_rmp}

The first-order expansion in the full $U(i \nu)$ leads to the 
following expression for the Green's function:
\begin{equation}
\label{fullU}
G_U(\tau) = \mathcal{G}_0(\tau) \exp \left( -\frac{1}{2} \iint_0^\beta \!\! \dd t \dd t^\prime \chi(t,t^\prime,\tau) U(t - t^\prime) \right).
\end{equation}
where $\mathcal{G}_0^{-1}(i \omega) = i \omega -\mu - \Delta(i \omega)$, and
$\chi(t,t^\prime,\tau) = \langle \mathcal{T} c_\tau c^\dag_0 N(t) N(t^\prime) \rangle_0 / \mathcal{G}_0(\tau)$ is
the connected part of the 
density-density correlator easily computed for the Gaussian $\mathcal{G}_0$ propagator.
$N(t)=\sum_\sigma \left(n_\sigma(t) - 1/2 \right)$ is the spin (and orbital) integrated density.
Note the exponential form in Eq.~\ref{fullU}, which comes from 
the first order cumulant expansion in $U$. 
According to the definition in Eq.~\ref{eq:ansatz},
the factor $F$ obtained in this way is given by the ratio $G_U(\tau)/G_{U_0}(\tau)$,
which reads
\begin{equation}
\label{first_order}
F_\textrm{\tiny GC}(\tau) = \exp \left( -\frac{1}{2} \iint_0^\beta \!\! \dd t \dd t^\prime \chi(t,t^\prime,\tau) {\bar U}(t - t^\prime) \right),
\end{equation}
with ${\bar U}(t)=U(t) - U_0\delta(t)$.
The above expression for the dynamic factor has the advantage to be very accurate at both small $U$, 
since the interaction is treated perturbatively, and strong coupling, as the density-density correlator 
will factor up by giving the exact dynamic atomic limit (see the Appendix). Incidentally, this is the reason
why the cumulant expansion is more effective than the regular perturbation theory for this case.

\subsection{Diagrammatic first-order expansion: instantaneous bold cumulants (IBC)}
\label{diagrammatic_deltaU}

One can go beyond the approximation for $F$ in Eq.~\ref{first_order} and compute the density-density correlator $\chi$
by using the interacting static $U_0$ Green's function $G_0(\tau)$ 
as propagator instead of the hybridized Green's function $\mathcal{G}_0(\tau)$.
This requires a correlated method, as the corresponding action is no longer Gaussian and the Wick theorem cannot be applied.
The QMC methods can compute $\chi$ directly in the interacting systems.\cite{antoine_rmp}
This development represents a consistent diagrammatic first-order expansion in ${\bar U}(i \nu)$
where the reference theory is the static $U_0$ model. The main difference with respect to the previous method reported in 
Subsec.~\ref{diagrammatic_fullU} stems from the fact that not only hybridization effects but also 
the impact of the instantaneous $U_0 \delta(t - t^\prime)$ interaction are included in $\chi$.
The cumulant expansion of the full Green's function in the retarded part ${\bar U}(i \nu)$ 
with $G_0(\tau)$ taken as the instantaneous bold propagator gives directly a factorized form of the type in Eq.~\ref{eq:ansatz}.
At the first order, it takes the expression in Eq.~\ref{fullU}
with $\mathcal{G}_0(\tau)$ replaced by $G_0(\tau)$ and the thermal averages $\langle \cdots \rangle_0$ computed
for the static interacting model, such that
$\boldsymbol{\chi}_{G_0}(t,t^\prime,\tau) = \langle \mathcal{T} c_\tau c^\dag_0 N(t) N(t^\prime) \rangle_{G_0}/G_0(\tau)$. 
It turns out that in this case the $F$ factor reads:
\begin{equation}
\label{ibc}
F_\textrm{\tiny IBC}(\tau) = \exp \left( -\frac{1}{2} \iint_0^\beta \!\! \dd t \dd t^\prime \boldsymbol{\chi}_{G_0}(t,t^\prime,\tau) {\bar U}(t - t^\prime) \right).
\end{equation}

In the following we report a comparison of the different approaches proposed in this work.
As common benchmark, we chose to compute the Green's function of the Anderson-Holstein impurity problem both in the symmetric 
and asymmetric cases, and check the BFA solutions against the numerically exact one provided by Rubtsov's CTQMC algorithm.

\section{Overview on the performance of the proposed factorization approximations}
\label{overview}

The factorization introduced in Eq.~\ref{eq:ansatz}
is an extremely useful Green's function ansatz to compute thermal and spectral properties 
coming from a generic retarded interaction in the multiband Anderson model,
and also the Hubbard model at the DMFT level.
The product in the time domain between 
a Bose factor $F$ embedding the dynamic properties of the interaction and
an auxiliary Green's function $G_0$ including instantaneous interactions and
low energy features,
leads to a deconvolution of the spectrum into low and high frequency contributions.
Its low frequency part, depending on $G_0$, can be easily obtained by available and
well developed ME methods, while the high energy features, as the
plasmon satellites, difficult or impossible to obtain by standard analytical continuation,
are directly given by the analytically known Bose factor $F$, 
which can be inverted in an accurate way by the Pad\'e approximants.

\begin{figure}[!ht]
\begin{center}
 \includegraphics[width=\columnwidth]{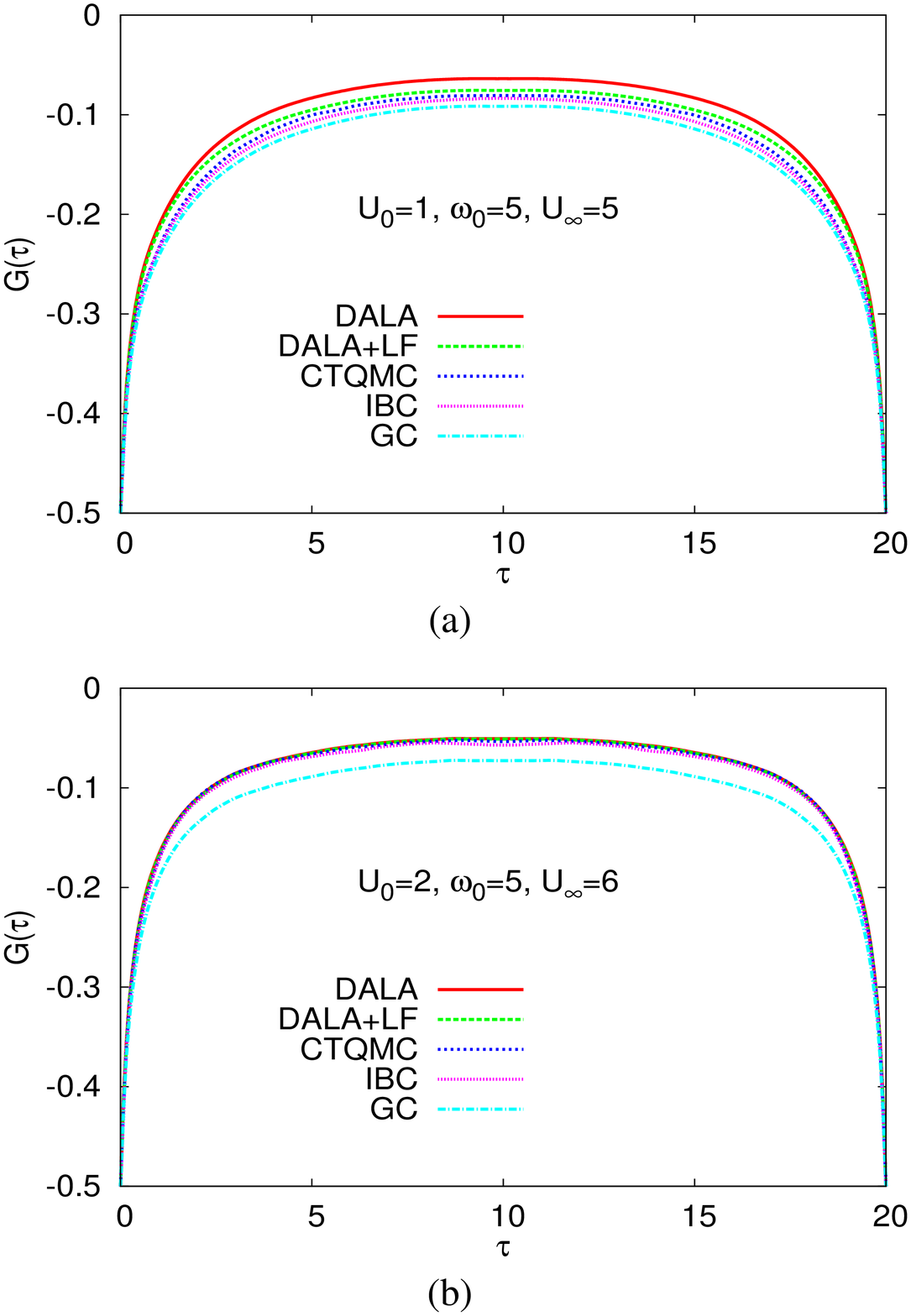}
\end{center}
\caption{\label{all_half_filling}
(color online) Symmetric Anderson model with retarded screened interaction as in Fig.~\ref{weaker}: 
$\omega_0=5$, $U_\infty-U_0=5$, $\beta=10$, for $\beta=20$ and two values of $U_0$,
(a) $U_0=1$; (b) $U_0=2$.
The numerically exact Rubtsov's CTQMC, DALA, DALA+LF, GC, and IBC Green's functions are reported.
}
\end{figure}

The Green's function Bose factor is a very general ansatz. However, the factor $F$ is not
exactly known in the generic case and needs some approximations. The most practical 
and physically insightful one
is borrowed from the dynamic atomic limit, where the factorization
is exactly given by a $G_0$ depending only on the instantaneous $U_0$ times
a factor which depends only on the retarded part ${\bar U}(i \omega)$.
The DALA proposed in Sec.~\ref{atomic} consists of keeping the atomic dynamic factor $F$
and taking the $G_0$ from the exact numerical solution of the instantaneous $U_0$ model.
As reported in Figs.~\ref{benchmark} and \ref{all_half_filling}(b), the DALA performs well
in the antiadiabatic regime when $\omega_0 > U_\infty - U_0$, and in 
the strong coupling regime, with a large $U_0$. 
The quality of the low energy part of the DALA Green's function worsens as the interaction becomes weaker
and $\omega_0$ gets smaller,
while its high energy tails are well reproduced even in the intermediate coupling.
The major failure of the DALA is the breaking of the Friedel sum rule, which is apparent
at low temperature (below $\beta=40$) or away from the antiadiabatic regime.
That is not surprising since the DALA is built upon the atomic limit.
In general, the DALA works 
when the high energy (unscreened) part of the retarded interaction is well separated (in frequency)
from the low energy (screened) part, which is the most common situation in the 
case of realistic Hamiltonians, with $U(i \omega)$ determined \emph{ab-initio} by the c-RPA approach\cite{crpa}.

\begin{figure}[!ht]
\begin{center}
 \includegraphics[width=\columnwidth]{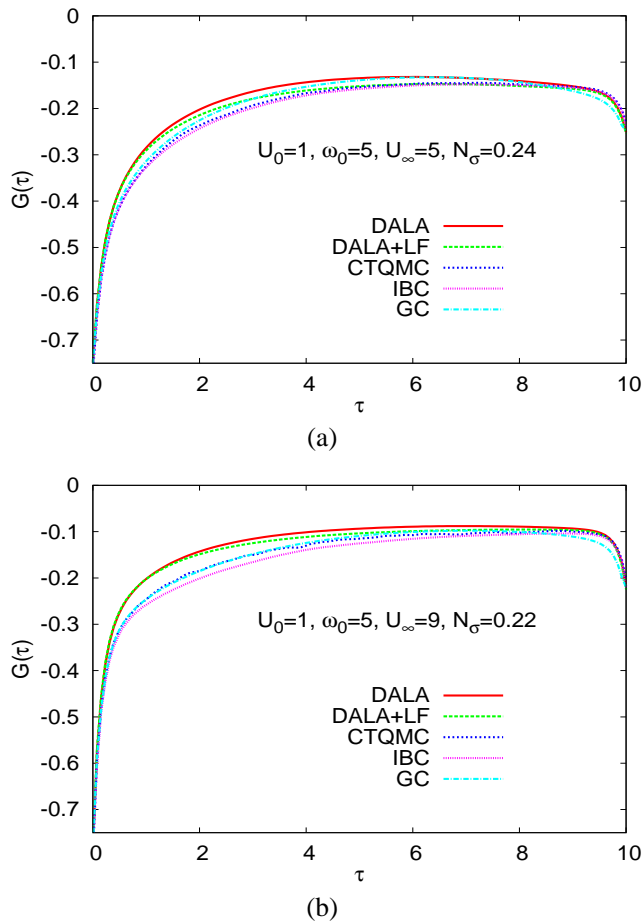}
\end{center}
\caption{\label{all_quarter_filling}
(color online) Asymmetric spin unpolarized Anderson model with retarded screened interaction as in Fig.~\ref{all_half_filling}: 
$\omega_0=5$, $U_\infty-U_0=5$, for $\beta=20$ and two values of $U_0$,
(a) $U_0=1$; (b) $U_0=2$.
In this case the number of particles $N_\sigma$ was set to be around the quarter filling.
The numerically exact Rubtsov's CTQMC, DALA, DALA+LF, GC, and IBC Green's functions are reported.
}
\end{figure}

An improvement upon the DALA is represented by the Lang-Firsov approximation, which fulfills
the Friedel sum rule and cures the low-energy low-temperature behavior. The way to 
incorporate the LF into the DALA factor, described in Subsec.~\ref{dalalf}, clearly improves the
low-energy features of the DALA Green's function, as one can see in Figs.~\ref{all_half_filling} and 
\ref{all_quarter_filling}, where the value of $G(\beta/2)$ yielded by the DALA+LF approach 
is closer to the exact numerical result in both the half-filling and quarter-filling Anderson-Holstein impurity models.
The DALA+LF factor provides an overall better agreement with the numerically exact Green's function
computed by the CTQMC method. The LF correction turns out to be important particularly
in the weak-intermediate coupling and at low temperature, while in other cases 
the simple DALA approximation is already good enough. For instance, in Fig.~\ref{all_half_filling}(b), 
when $U_0$ is strong and the temperature not-so-low ($\beta=10$), 
the DALA Green's function coincides with the DALA+LF one, and both are on the top 
of the exact numerical solution. A common limitation of the DALA and DALA+LF approximations is that 
they become inaccurate when $\omega_0 < U_\infty-U_0$, while the work very well in the antiadiabatic regime.

A different route to determine $F$ is built on diagrammatic expansion techniques.
We have proven that the cumulant expansion of the Green's function up to the first order in the interaction
provides the exact connection to the dynamic atomic limit (see Appendix). Indeed, the $F$ factor computed 
in this way (see Eq.~\ref{first_order}) has the exact atomic behavior 
in the zero hybridization limit. We propose two flavors for the cumulant expansion.
In both cases one has to evaluate the density-density correlator $\langle \mathcal{T} c_\tau c^\dag_0 N(t) N(t^\prime) \rangle_0$.
In the first one (the Gaussian cumulants or ``GC'') the thermal average is computed with a Gaussian action based on the hybridized 
non-interacting Green's function, while in the other one (the instantaneous bold cumulants or ``IBC'') the brakets are computed with
the instantaneous $U_0$ Green's function.
The GC works well in the weak interacting regime (small $U_0$) and it worsens as $U_0$ is getting larger,
as shown in Fig.~\ref{all_half_filling}. Away from half-filling, the GC Green's function is reliable in the
``empty'' part of the spectrum (for $\tau < \beta/2$), and where the dynamic part is more relevant in setting the 
tails of the Green's function. On the other hand, the IBC performs well everywhere in all cases when $\omega_0$ is large,
as seen in Figs.~\ref{all_half_filling} and \ref{all_quarter_filling},
and it is supposed to work quite well even away from the antiadiabatic regime, since it retains the feedback of the 
instantaneous interaction on the dynamic part via the $\chi$ correlator.
The price to pay in the latter case is the statistical uncertainty of $F_\textrm{IBC}$,
as $\chi$ must be evaluated by means of Monte Carlo techniques, in contrast to the GC approach where $\chi$
is known up to machine precision thanks to Wick's theorem. This could lead to some inaccuracy 
in the analytical continuation of the IBC Bose factor.

If the goal is to compute the spectral representation of a Green's
function with retarded interactions, we found that the most effective
way is to use the Bose factor taken from the dynamic atomic limit.
As already reported in Sec.~\ref{ansatz}, one has to define an auxiliary
Green's function $G_\textrm{aux}(\tau)$ as
$G(\tau)/F_\textrm{DALA}(\tau)$, and then use
Eq.~\ref{spectral_conv} to compute the spectrum.
The DALA factor is the simplest to invert, and its bosonic spectral
representation is the ``physical'' density of plasmonic (or phononic) modes, whereas
the other factors beyond the DALA include also some low energy 
contributions. Therefore, 
the DALA factor is the most recommended in the assisted analytical
continuation.

\section{Application to the single-band Hubbard-Holstein model}
\label{results}
\label{half_filled_bethe}

In the following, we are going to present applications
where our BFA
approach is used as solver
of the Anderson impurity problem resulting from the
DMFT self-consistency conditions. Therefore, 
in contrast to what has been shown in the methodological Sections, 
the following results are the converged
solutions of the DMFT equations for the full lattice Hamiltonian, which incorporates
the retarded $U(i \nu)$ as the on-site interaction.

\begin{figure}
\begin{center}
 \includegraphics[width=\columnwidth]{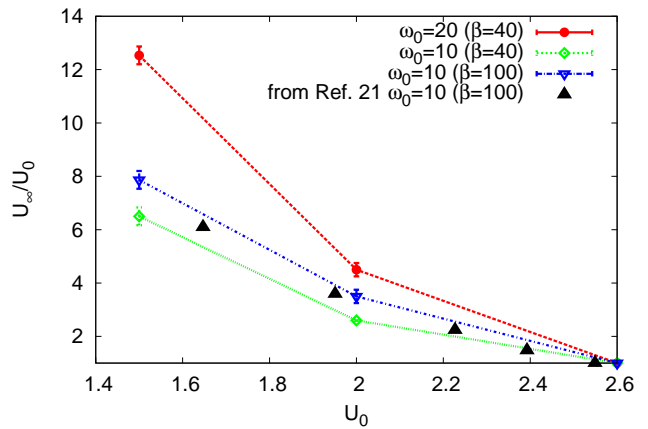}
\end{center}
\caption{\label{critical_U}
(color online) Critical $U_\textrm{c2}$ line of the Mott transition 
calculated at $\beta=40$ and half-filling on a Bethe lattice for the
screened instantaneous interaction $U_0$ as a function of $U_\infty/U_0$ and $\omega_0=10,20$ 
(green diamonds and red circles respectively).
The critical values are obtained from the Green's function computed by the DALA+LF ansatz.
The behavior of the critical screened $U$ is in agreement with
the one computed by Werner and Millis with their CTQMC algorithm at $\omega_0=10$ 
and $\beta=100$ (black upper triangles). Indeed, the critical line
calculated by means of our DALA+LF approach (blue lower triangles)
is on top of their CTQMC points\cite{werner_millis}.
This highlights the accuracy of our approximated DALA+LF
method if compared to the exact numerical result.
Note the quite strong temperature dependence 
of the critical $U_\textrm{c2}$ in going from $\beta=40$ to $\beta=100$. 
This is a quite interesting effect which deserves further analysis.
}
\end{figure}

The first application is on
the single band half-filled Hubbard-Holstein model solved by the DMFT 
on the Bethe lattice. 
To study the impact of the retarded screened interaction on the Mott transition,
we analyze the half-filled model
at $\beta=40$ with different $U_0$, $U_\infty$, and $\omega_0$ parameters.
We choose to work always with $U_\infty > U_0$, a ``physical'' condition
which states that the unscreened $U$ is larger than the screened one.
The screening frequency is taken such that we are in the antiadiabatic regime,
and we use our various Bose factor approximations to predict 
the critical value of $U_\infty$ for the Mott transition, once the other parameters
are fixed, such that $\omega_0=10$ and $U_0=2$. 
As reported in literature,\cite{hubbard_holstein,hubbard_holstein2} 
the metal-Mott insulator transition
in the Hubbard-Holstein model is first order.
In Tab.~\ref{table} we present our lower and upper critical values 
obtained for $\beta=40$ by means of the BFA at various levels of approximation.
\begin{table}[!ht]
\caption{Upper and lower critical $U_\infty$ values obtained at $\beta=40$ 
by different BFA methods for the Mott transition in the single-band
Hubbard-Holstein model on the Bethe lattice at half-filling, once $U_0(=2)$ and $\omega_0(=10)$ are fixed.
The values are expressed in the half-bandwidth units. 
By our BFA we always get a first order phase transition, with a hysteresis effect, 
in agreement with what reported in literature\cite{hubbard_holstein2}.}
\label{table}
\begin{ruledtabular}
\begin{tabular}{|l d d|} 
   \makebox[0pt][l]{method}  & \makebox[0pt][c]{$U^\textrm{c1}_\infty$}  &  \makebox[0pt][c]{$U^\textrm{c2}_\infty$} \\
\hline
GC      &   12.3  &   12.5 \\
DALA    &    8.2  &    8.4 \\
DALA+LF &    4.9  &    5.2 \\
\end{tabular}
\end{ruledtabular}
\end{table}
As one can see, the actual values at criticality depend quite strongly on
the approximation used, with the DALA+LF giving the results aligned with the numerically exact CTQMC
method by Werner and Millis\cite{werner_millis} (see Fig.~\ref{critical_U}). 
In order to have a close comparison to the data published in Ref.~\onlinecite{werner_millis},
we carried out some calculations also for $\beta=100$. The agreement between the
DALA+LF and the Monte Carlo predictions is quite remarkable.
This highlights again the importance of the LF correction at low energy 
in order to accurately predict the physical properties.

Fig.~\ref{critical_U} summarizes the results for the upper critical line
for different sets of $\omega_0$ and $U_0$, whose values are
taken not so far from the critical $U_\textrm{c2} \approx 2.6$ 
of the static model computed at $\beta=40$\cite{mott_transition_dmft,mott_transition_dmft_antoine}.
It is clear that the fully retarded model with $U_\infty > U_0$
is more correlated than
the static one with the same instantaneous $U_0$.
The Mott transition happens at values of screened $U_0$ \emph{lower} than 
the critical static $U_\textrm{c2}$ for any finite $\omega_0$. 
The dependence of the critical parameters on $\omega_0$ is also clear.
At fixed screened $U_0$, a smaller $\omega_0$ corresponds to 
a smaller unscreened $U_\infty$ at which the Mott transition is reached.  
Indeed, if the frequency of the screening plasmon is closer to 
the Fermi level, it is easier for the unscreened part to induce
the Mott transition at low energy. The same conclusions were reached 
by Werner and Millis by means of their numerically exact CTQMC algorithm.
This shows the impact of the dynamic screening features on
the low energy properties of the model. 
In order to get the same effective low energy parameters, 
the ``effective'' instantaneous
$U_0$ can be up to $20-25\%$ larger than the true screened value even
for plasmon frequencies $\omega_0 (\gg U_0, \gg D)$ in the antiadiabatic regime.

\begin{figure}
\begin{center}
 \includegraphics[width=\columnwidth]{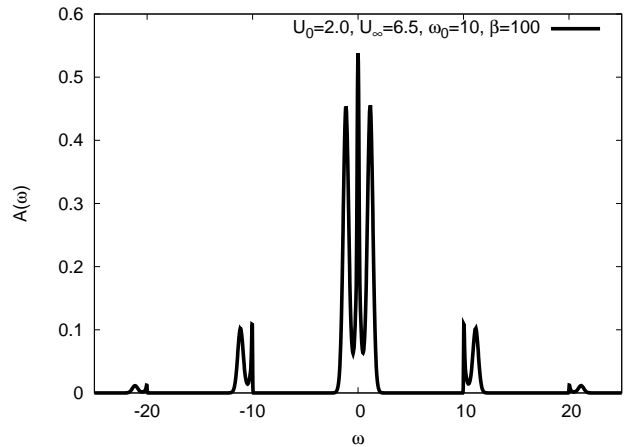}
\end{center}
\caption{\label{spectral_function_critical_U}
Spectral representation of the Green's function for a Holstein-Hubbard model with 
$\omega_0=10$, $U_0=2$, $U_\infty=6.5$ and $\beta=100$, close to the blue Mott transition line
of Fig.~\ref{critical_U}. The analytic continuation has been assisted by the DALA Bose factor,
which allows one to describe accuratley the plasmon satellites centered around frequences multiple of $\omega_0$.
}
\end{figure}

In Fig.~\ref{spectral_function_critical_U} we report the spectral representation of a 
half-filled Holstein-Hubbard model with $\omega_0=10$, $U_0=2$, $U_\infty=6.5$, and temperature $\beta=100$,
quite close to the Mott transition. The spectral function has been obtained with the help of the DALA factor
by the method in Sec.~\ref{ansatz}. It describes with high accuracy 
not only the low-energy features but also the plasmon satellites in the anti-adiabatic regime, typical of
realistic materials.

\section{Realistic application to $\mathbf{SrVO_3}$}
\label{srvo3}

Our second application 
is on a realistic Hamiltonian for \srvo3, a very well studied 
material which represents a benchmark for theories describing
strongly correlated compounds.
\srvo3 
is the prototype of 
a correlated metal where the many-body treatment of 
correlation in the $d$ manyfold is important to
explain 
the spectral properties\cite{srvo3_exp4_intro}. 
Therefore, it 
has been the subject of intensive studies\cite{liebsch_dmft,pavarini_dmft,pavarini_dmft2,nekrasov_dmft,nekrasov_dmft2,lda_dmft_lapw} 
applying DMFT 
in the context of realistic strongly correlated Hamiltonians.
Indeed, its band structure is relatively simple due to its undistorted 
perovskite 
structure,
resulting in the occupation of one electron in three 
degenerate $t_{2g}$ bands crossing the Fermi level.
The $p$ Oxygen ligands are quite well separated from the $d$ levels, 
such that the definition of a low-energy \t2g-Hamiltonian
is unambiguous. 
Thus, \srvo3 has been the testing case for many new DFT-DMFT 
implementations.\cite{full_orbital_anisimov_dmft,lda_dmft_wannier,
lda_dmft_pw_projected,lda_dmft_pseudo_pw,lda_dmft_lapw}
On the other hand, \srvo3 has been the subject of intensive experimental 
activity, with magnetic, electrical, optical measurements
\cite{srvo3_exp1_resistivity_susceptibility,srvo3_exp5_resistivity_thermodynamic_magnetic,srvo3_exp6_optical},
and by means of photoemission spectroscopy (PES)\cite{srvo3_exp2_pes,srvo3_exp3_early,srvo3_exp7_pes_bulksurface,srvo3_exp8_pes_dmft} 
and angled resolved PES (ARPES)
\cite{srvo3_exp9_arpes,srvo3_exp10_arpes_tightbinding,srvo3_exp11_arpes_bulksurface,srvo3_exp12_arpes_thin_dmft,srvo3_exp13_latest_arpes}.

Here, we consider a model where only the
$t_{2g}$ electron is retained, and all the others contribute
to screen the local lattice interaction. The DFT band structure
has been calculated with the linear muffin-tin
orbital (LMTO) framework in the atomic sphere approximation (ASA),
which allows one to work with a native $d$-projected and localized 
orbital representation.
The realistic retarded $U$ for this compound has been computed in Ref.~\onlinecite{crpa2_srvo3_lmto}
based on the c-RPA construction.

The low-energy Hamiltonian we are going to work with 
consists of the LDA \t2g-Hamiltonian, the plasmonic
part giving rise to the dynamical screening and the
following static interaction Hamiltonian:
\begin{eqnarray}
\label{hubmod}
H_{U_0} && = U\sum_mn_{m\uparrow} n_{m\downarrow}+
\frac{U'}{2} \mathop{\sum_{mm'\sigma}}_{m\ne m'} n_{m\sigma}n_{m'\bar{\sigma}}\nonumber\\
&& + \frac{U''}{2}\mathop{\sum_{mm'\sigma}}_{m\ne m'} n_{m\sigma}n_{m'\sigma} 
\end{eqnarray}
where $n_{m\sigma}$=$d_{m\sigma}^{\dag}d_{m\sigma}^{\hfill}$ is the usual density operator, 
with $m$,$\sigma$ denoting orbital and spin indexes, and 
$d_{m\sigma}^{\dag}$, $d_{m\sigma}^{\hfill}$ representing the 
$t_\textrm{2g}$ localized orbitals.
For the \t2g-orbitals of SrVO$_3$, $U'=U -2J$ and $U''= U- 3J$,
with the screened value of the interaction $U_0=3.6$ eV, 
and the Hund's coupling $J=0.68$ eV.
The additional retarded interaction 
which is included in our model, couples to the total charge
of the system, as described by Eq.~\ref{generic_interaction}.
There is no need for an explicit double-counting term, since
such a correction is absorbed into the effective chemical
potential fixing the particle number to one.

\begin{figure}
\begin{center}
 \includegraphics[width=\columnwidth]{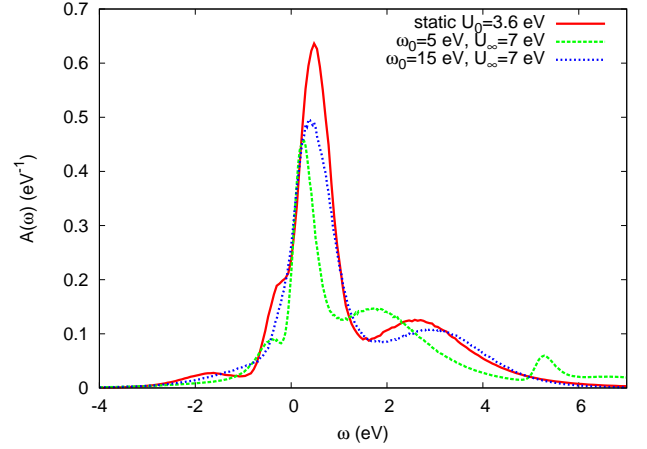}
\end{center}
\caption{\label{static_vs_dynamic}
(color online) Spectral function of 
one electron in 3 $t_{2g}$ bands for the \srvo3
obtained by the DFT in the local density approximation.
The DMFT calculations are done at $\beta=10$ eV$^{-1}$ with the DALA Bose factor solver
for the retarded $U$ and the Hirsch-Fye QMC solver for the instantaneous interaction.
Three types of interactions are reported:
the purely static one, with $U_0=3.6$ eV, a dynamic Holstein one with the same 
instantaneous interaction but with the unscreened $U_\infty=7$ eV and 
a plasmon frequency $\omega_0=5$ eV, another Holstein interaction
with the same $U_0$ and $U_\infty$ but different $\omega_0(=15 \textrm{ eV})$.
Note the displacement of the upper Hubbard shoulder shifted 
at lower energy by the presence of the plasmon at $\omega_0=5$ eV.
}
\end{figure}

To study the impact of  $U_\textrm{retarded}$ on the low energy properties 
of the model, we took into account different Holstein single-plasmon $U(i \nu)$'s, and compared them
with the corresponding static model.
We used our BFA approach in its simplest DALA formulation, and
computed the spectrum at $\beta=10$ and $\beta=20$ eV$^{-1}$.
Indeed, it turns out that at those temperatures the LF correction is irrelevant, and so
the DALA performs here at best even in terms of efficiency. 
In Fig.~\ref{static_vs_dynamic}, we report the spectra computed 
with the instantaneous $U_0(=3.6 \textrm{ eV})$ interaction, and two retarded 
interactions with the same unscreened $U_\infty(=7 \textrm{ eV})$, but different
screening $\omega_0$'s (one at 5 eV, the other at 15 eV).
The static model has a quasiparticle peak at the Fermi level typical of strongly correlated 
compounds, with a lower and upper shoulders reminiscent of the lower and upper Hubbard bands.
In the dynamic model, the effect of the 15 eV plasmon is to
renormalize the quasiparticle spectral weight, transferred at higher energies, while the
shape of the low energy spectrum is almost unchanged. The effect of the
5 eV plasmon, closer to the low energy sector, is more remarkable.
Beside a stronger spectral weight reduction, there is a shift of the
upper Hubbard shoulder to lower energies than in the corresponding static model.
This effect is certainly due to the interplay between the plasmon satellite
at 5 eV, visible in Fig.~\ref{static_vs_dynamic}, and the low energy features of the spectrum.

\begin{figure}
\begin{center}
 \includegraphics[width=\columnwidth]{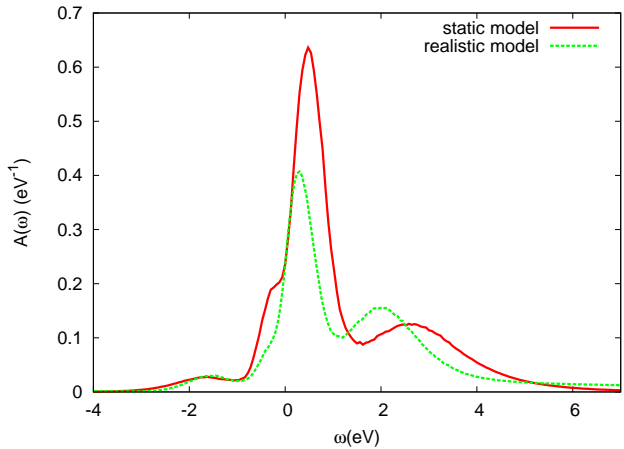}
\end{center}
\caption{\label{static_vs_realistic}
(color online) Spectral function at $\beta=10$ eV$^{-1}$ of 
one electron in 3 $t_{2g}$ bands for the \srvo3
obtained by the DFT in the local density approximation.
The DMFT calculations are carried out with the DALA Bose factor solver
for the retarded $U$ and the Hirsch-Fye QMC solver for the instantaneous interaction.
The result of the realistic Green's function is reported,
showing a more correlated behavior than the corresponding static model 
with the same instantaneous $U_0 (= 3.6 \textrm{ eV})$. The quasiparticle peak is smaller
than the corresponding static model, with a spectral weight transfer at higher energies.
}
\end{figure}

The realistic retarded $U(i \nu)$ is characterized by two main
screening frequencies at 5 and 15 eV\cite{crpa2_srvo3_lmto,crpa3_srvo3_wannier},
while the unscreened value of $U$ is 16 eV, much larger than the one
in the models considered so far. By applying our Bose factor ansatz 
to this problem, we found the spectral function in 
Fig.~\ref{static_vs_realistic}, reported together with the
instantaneous interaction with the same screened $U_0(=3.6 \textrm{ eV})$.
Also here we note the quasiparticle weight reduction, and 
the shift of the upper Hubbard shoulder to lower frequencies, 
as in the model $U$ analyzed before,
while the lower Hubbard band is almost unchanged with respect of the static model. 
Its maximum corresponds to the position found in PES measurements,
and well documented in previous DMFT studies on \srvo3.
The weight reduction of the quasiparticle peak 
coincides with a smaller value of $Z=1/\left(1- \left. \partial \Sigma^{''}(i \omega) / \partial \omega \right|_{\omega=0}\right)$,
namely a larger value of the effective mass $m^*$.
In particular, for the realistic dynamic model we found 
a $Z \approx 0.5$, which gives an effective mass renormalized by 2 with respect to the DFT band structure, 
while for the corresponding static model we obtained a value of $Z \approx 0.7$.
Recent ARPES data yielded an effective mass 
$m^* \approx 2 m_0$\cite{srvo3_exp9_arpes,srvo3_exp11_arpes_bulksurface,srvo3_exp13_latest_arpes}, 
which is in a good agreement with our findings for the realistic
retarded interaction. On the other hand, the static model with the same
screened $U_0$ underestimates the correlation by a factor of $1.4$, 
and so the value of $m^*$. In the literature, a static model 
with a larger instantaneous $U_0(\approx 5 \textrm{ eV})$  has been used to 
find the experimental mass renormalization\cite{pavarini_dmft,pavarini_dmft2,nekrasov_dmft,nekrasov_dmft2}.
Such a larger value of $U$ could be justified by the constrained LDA method\cite{constrained_lda_first,constrained_lda}
used to determine \emph{a priori} the on-site interaction, but known to overestimate its strength. The difficulty here
is to reproduce by the same model both the effective mass 
and the position of the lower Hubbard band, which turns out to be 
shifted at lower energies ($\approx -2$ eV) by a stronger $U_0$. 
This is the reason why in Ref.~\onlinecite{lda_dmft_wannier}
the authors made the choice to work with $U_0=4$ eV,
a slightly weaker effective static interaction which gives the lower Hubbard band correctly
peaked at $-1.5$ eV. Some cluster calculations give the same position for the lower Hubbard band,
with the interpretation that its peak depends on the hybridization and screening 
properties provided by the ligands\cite{cluster_srvo3,cluster_srvo3_second}. 
With our dynamically screened model we describe correctly both the 
mass renormalization ($2 m_0$) and the lower Hubbard band position peaked at $-1.5$ eV, 
as apparent in Fig.~\ref{static_vs_realistic}.
This analysis highlights the importance of including
the proper \emph{retarded} interaction, to have a
reliable and fully \emph{ab-initio} description of the correlation
in these materials.

We note however, that the present description includes only
the \t2g orbitals, so that the above conclusions are valid
in an energy range where no other orbital contributions
are present.

\section{Conclusions}
\label{conclusions}

We introduced a factorized form $G=FG_0$ for the Green's function of the Anderson model 
with generic retarded (dynamic) interaction, dubbed ``Bose factor ansatz''. 
We proposed various approximations for the Green's function Bose factor $F$,
the most practical and effective one borrowed directly from 
the dynamic atomic limit (DALA), whose form is analytically known.
The DALA provides an accurate way to compute the Green's function
in the antiadiabatic limit \emph{and} evaluate the spectral properties
in the full frequency range by means of an improved analytical continuation method.
In practice, the inversion from the imaginary to the real frequency domain
is assisted by the factor $F$ which retains the main information on the 
position and strength of the plasmon satellites, and enters in
the analytical continuation as a convolution with the instantaneous part $G_0$.
We carefully analyzed the pros and cons of the DALA, and found various ways 
to improve $F$, by either using the Lang-Firsov approximation
at low-frequency, or resorting to diagrammatic techniques.
Finally, we applied our approach to lattice problems in the 
context of the DMFT formalism. We took into account
the Hubbard-Holstein Hamiltonian at half-filling and 
on the Bethe lattice, and we studied the Mott transition
in the spirit of looking at the retarded interaction as
resulting from ``realistic screening'' of the bare $U$ by a single plasmon.
The second application has been for the realistic \srvo3 Hamiltonian,
where the $t_\textrm{2g}$ electrons interact via a retarded 
on-site $U$ previously determined \emph{ab-initio} at the c-RPA level\cite{crpa}.
In both cases, it turns out that it is important 
to retain the retarded features of the local interaction 
resulting from the dynamic screening in order to have
a reliable \emph{ab-intio} description of 
correlated materials. Also, our approach could be useful
to determine whether some spectral signatures at intermediate energy ($\approx 10$ eV)
seen in a broad class of correlated materials\cite{chainani} 
come from a bulk dynamic screening.
In perspective, more work has to be done theoretically to rationalize
the effects of the screening on both the ground and excited states properties
of correlated compounds.
Morever, by means of the same formalism, one can study Jan-Teller models to
describe the impact of the electron-phonon coupling to the spectral
function of the distorted compound.
Last but not least, dealing with a frequency dependent $U$ is an essential step
toward the implementation of the GW+DMFT framework\cite{silke_gwdmft}, where 
the screening resulting from the GW polarization has to be included consistently in
the low-energy correlated model solved at the DMFT level. 
Therefore, behind this work 
there are important
experimental, theoretical and methodological implications that one can now 
start taking into account.

\appendix*
\section{From the first-order cumulant expansion to the dynamic atomic limit}
\label{appendix}

In this appendix we prove that the Bose factor $F$ reported 
in Eq.~\ref{first_order}
fulfills the exact atomic limit, i.e. it equals the one in 
Eq.~\ref{eq:dala} for $\Delta=0$. 
This is a non trivial property, which guarantees the GC approximation 
to provide a reasonable description of the insulating phase 
(and the metal-to-insulator transition) 
in the DMFT framework, where the hybridization function of the 
Anderson impurity 
assumes a crucial frequency dependence in order to represent 
the coupling with the self-consistent bath.
In the strong coupling case, the hybridization 
$\Delta(i \omega_n)$ goes to zero for small $\omega_n$.
Thus, in the DMFT language, this phase is mapped into an 
Anderson impurity problem
close to the atomic limit, and its accurate solution is 
required around $\Delta=0$.

We start by noting that the density-density correlator 
$\chi$ factorizes in the atomic limit, as
\begin{equation}
\label{atomic_chi1chi1}
\chi(t,t^\prime,\tau) {\underset{\Delta=0}{=}} \chi^1(t,\tau) \chi^1(t^\prime,\tau)
\end{equation}
where $\chi^1(t,\tau)=\langle \mathcal{T} c_\tau c^\dag_0 N(t) \rangle_0/\mathcal{G}_0(\tau)$.
In the most general case, $\chi$ can be resolved into its 
spin and orbital components,
by defining $\chi_{\sigma \sigma^\prime}(t,t^\prime,\tau) = 
\langle \mathcal{T} c_\tau c^\dag_0 N_\sigma(t) N_{\sigma^\prime}(t^\prime) \rangle_0 / \mathcal{G}_0(\tau)$,
with $N_\sigma(t)=n_\sigma(t) - 1/2$. 
The factorization in Eq.~\ref{atomic_chi1chi1}
holds also for the spin resolved quantities:
$\chi_{\sigma\sigma^\prime}=\chi^1_\sigma \chi^1_{\sigma^\prime}$.
Once the Wick theorem is applied and the former correlators are written
in terms of the $\mathcal{G}_0(\tau)$'s, it is straightforward 
to prove the
spin resolved identity and consequently the integrated one 
of Eq.~\ref{atomic_chi1chi1},
by using the atomic limit expression for $\mathcal{G}_0(\tau)$.
For instance, in the atomic limit and for a generic $\mu$, 
the connected part of $\chi^1$ reads:
\begin{equation}
\label{atomic_chi1}
\chi^1(t,\tau) = \left \{
\begin{array}{ll}
-e^{-\mu \beta}/\left(1 + e^{-\mu \beta}\right)    & \textrm{for $\tau < t < \beta$} \\
1/\left(1+e^{-\mu \beta}\right)                   & \textrm{for $0 \le t \le \tau$}
\end{array} 
\right. .
\end{equation}

Now we use a Hubbard-Stratonovich (HS) transformation to rewrite the
$F_\textrm{GC}$ factor in a form which includes $\chi^1$ in linear terms only.
We obtain the following identity:
\begin{eqnarray}
\label{hubbard_stratonovich}
\exp \left( -\frac{1}{2} \iint_0^\beta  \!\! \dd t \dd t^\prime \chi^1(t,\tau) {\bar U}(t - t^\prime) \chi^1(t^\prime,\tau) \right) & = \nonumber \\ 
\int {\mathcal D}\phi \exp \left( -i \int_0^\beta  \!\! \dd t \chi^1(t,\tau) \phi(t) \right.  & \nonumber \\
\left. - \frac{1}{2} \iint_0^\beta  \!\! \dd t \dd t^\prime \phi(t) {\bar U}^{-1}(t-t^\prime) \phi(t^\prime) \right), &  
\end{eqnarray}
where $\int {\mathcal D}\phi$ is the functional integral over the complex HS field $\phi(t)$.
By exploiting the atomic limit expression of $\chi^1$ in Eq.~\ref{atomic_chi1}, one can compute the integral involving 
the product of $\phi$ and $\chi^1$, which gives the result:
\begin{eqnarray}
\label{linear_chi1}
\int_0^\beta  \!\! \dd t \chi^1(t,\tau) \phi(t) & = & \left( \frac{\tau}{\beta} - \frac{e^{-\mu\beta}}{1+e^{-\mu\beta}}\right)\phi_0  \nonumber \\ 
                                       & +  & \frac{i}{\beta} \sum_{n \ne 0} \frac{\phi_n}{\nu_n} \left( e^{-i \nu_n \tau} - 1 \right),
\end{eqnarray}
with $\phi_n=\int_0^\beta \dd t \phi(t) e^{i \nu_n t}$ the Fourier components of the HS field.
By integrating the functional integral in Eq.~\ref{hubbard_stratonovich} in the Fourier space, one gets the final 
expression for the exponent of the GC factor:
\begin{eqnarray}
\label{semifinal}
 -\frac{1}{2} \iint_0^\beta \!\! \dd t \dd t^\prime \chi(t,t^\prime,\tau) {\bar U}(t - t^\prime)  {\underset{\Delta=0}{=}} &  \nonumber \\  
- \frac{1}{2} \beta {\bar U}(i \nu_0) \left(\frac{e^{-\mu\beta}}{1 + e^{-\mu\beta}}- \frac{\tau}{\beta} \right)^2 + & \nonumber \\
 \frac{1}{\beta} \sum_{n \ne 0} \frac{{\bar U}(i \nu_n)}{\nu_n^2} \left( e^{i \nu_n \tau} - 1 \right) &
\end{eqnarray}
We note here that ${\bar U}$ vanishes in the static limit, as
${\bar U}(i \nu_n) = U(i \nu_n) - U_0$. Therefore, the first term in the right-hand side of the above equation vanishes as well.
We are left with the last term in Eq.~\ref{semifinal}, which is exactly equal to the exponent of the DALA factor.
Thus, we have proven that
\begin{equation}
F_\textrm{GC}(\tau)  {\underset{\Delta=0}{=}} F_\textrm{DALA}(\tau),
\end{equation}
for an arbitrary $\mu$ and ${\bar U}$.

To conclude, the fact that the first-order cumulant expansion fulfills the exact dynamic atomic limit
justifies the use of cumulants instead of the standard first-order developements, 
and validates also the cumulant of the instantaneous bold factor in Eq.~\ref{ibc}.

\begin{acknowledgments}
This work was supported by IDRIS/GENCI under
grant number 101393, and by the French ANR under
project SURMOTT.
We thank Ferdi Aryasetiawan, Alexander I. Lichtenstein,
Andrew J. Millis, Takashi Miyake, Jan M. Tomczak
and Philipp Werner
for stimulating discussions.
\end{acknowledgments}

\end{document}